\documentclass[11pt,twoside]{article}

\usepackage{asp_MSF07}
\usepackage{epsf}
\usepackage{graphicx}
\usepackage{lscape}
\usepackage{natbib}

\newcommand{\sm}[1]{\mbox{{\scriptsize #1}}}

\newcommand{\simge} {\,{}^>_{\sim}\,}

\newcommand{\be}{\begin{equation}}
\newcommand{\ee}{\end{equation}}
\newcommand{\bea}{\begin{eqnarray}}
\newcommand{\eea}{\end{eqnarray}}
\newcommand{\bdm}{\begin{displaymath}}
\newcommand{\edm}{\end{displaymath}}
\newcommand{\bef}{\begin{figure}}
\newcommand{\eef}{\end{figure}}
\newcommand{\befone}{
  \begin{figure*}
  \centering
  \begin{minipage}{\textwidth}
  }
\newcommand{\eefone}{\end{minipage}\end{figure*}}

\newcommand{\cm}{\mbox{cm}}
\newcommand{\m}{\mbox{m}}
\newcommand{\km}{\mbox{km}}
\newcommand{\AU}{\mbox{AU}}

\newcommand{\g}{\mbox{g}}

\newcommand{\Msol}{\mbox{M$_{\sun}$}}
\newcommand{\pc}{\mbox{pc}}

\newcommand{\K}{\mbox{K}}
\newcommand{\yr}{\mbox{yr}}
\newcommand{\ys}{\mbox{yrs}}

\newcommand{\Ma}{{\cal M}}

\def\eps@scaling{0.95}

\def\showone#1{
  \centering
  \leavevmode
  \epsfxsize=\eps@scaling\linewidth
  \epsfbox{#1.eps}
}

\def\epstwo@scaling{0.44}

\def\showtwo#1#2{
  \centering
  \leavevmode
  \epsfxsize=\epstwo@scaling\linewidth
  \epsfbox{#1.eps}
  \epsfxsize=\epstwo@scaling\linewidth
  \epsfbox{#2.eps}
}

\begin{document}
\title{Filaments, Collapse and Outflows in Massive Star Formation}   
\author{Robi Banerjee$^1$ and Ralph E.~Pudritz$^2$}   
\affil{$^1$ Institute of Theoretical Astrophysics, University of
Heidelberg, Albert-Ueberle-Str. 2, 69120 Heidelberg, Germany \\
$^2$ Kavli Institute for Theoretical Physics, University of California,
Santa Barbara, CA 93106-4030}    

\begin{abstract} 
We present results from our numerical simulations of collapsing
massive molecular cloud cores. These numerical calculations show that
massive stars assemble quickly with mass accretion rates exceeding
$10^{-3} \, \Msol \, \yr^{-1}$ and confirm that the mass accretion
during the collapsing phase is much more efficient than predicted by
selfsimilar collapse solutions, $\dot{M} \sim c^3/G$. We find that
during protostellar assembly out of a non-turbulent core, the mass
accretion reaches $20 - 100 c^3/G$. Furthermore, we explore the
self-consistent structure of bipolar outflows that are produced in our
three dimensional magnetized collapse simulations. These outflows
produce cavities out of which radiation pressure can be released,
thereby reducing the limitations on the final mass of massive stars
formed by gravitational collapse.

Additional enhancement of the mass accretion rate comes from accretion
along filaments that are built up by supersonic turbulent
motions. Our numerical calculations of collapsing turbulent cores
result in mass accretion rates as high as $10^{-2} \, \Msol \,
\yr^{-1}$. 
\end{abstract}



\section{Introduction}

Our understanding of how massive stars form, while still far from
complete has recently made significant strides~\citep[see also Harold
Yorke's contribution in this proceedings and the review
by][]{ZinneckerYorke07}.  Low mass stars accrete the bulk of their
mass through their circumstellar disks before nuclear burning turns on
(e.g., \citealt{Shu87}).  Massive stars on the other hand, have
Kelvin-Helmholtz time scales that are smaller than the dynamical time
so that young massive stars start to burn their nuclear fuels while
still accreting the surrounding gas \citep[e.g.,][and references
herein]{Yorke02b, Yorke04}.  Infall, or flow of gas through a
surrounding disk, therefore faces a major obstacle in the form of the
radiative pressure that such massive stars will produce as they are
still forming.

Early spherical accretion models suggest that the resulting radiation
pressure could limit the final mass of the star to $\sim 40 \, \Msol$
if the accretion rate is not high enough ($\simge 10^{-3} \, \Msol \,
\yr^{-1}$) \citep{Kahn74, Wolfire87}. More recent two dimensional
simulations by \citet{Yorke02} showed that the limitations on the
final mass of the massive star can be relaxed if accretion through the
protostellar disk is included in models of massive star
formation. Still, even this particular result yields mass limits of
$\sim 43 \, \Msol$. \citet{Krumholz05} argued that the effects of
radiation pressure are limited by the escape of radiation through an
outflow cavity.  These results were based on simulations, using
Monte-Carlo-diffusion radiative transfer models where the outflow
cavities are parameterized by varying opening angles.

Alternatively, massive stars could form through coalescence of
intermediate mass stars~\citep[][]{Bonnell98}. This formation process
would be starkly different from the formation of low mass stars which
assemble quickly through accretion of the molecular gas. So far,
observations of the intermediate state of massive star formation are
rare and difficult to obtain. Nevertheless, a few massive objects
which show evidence for an ongoing accretion process are known by now
\citep[][]{Chini04, Patel05, Chini06, Beltran06}.

Here we summarize our recent effort studying the collapse of massive
cloud cores which final state will be one or more massive stars. These
cores are either modeled as supercritical hydrostatic spheres or
taken from simulations of supersonic turbulence in which the unstable
cores are formed in shock compressions. The numerical setups and
the technical details which also includes our description of various
cooling and heating processes can be found in~\cite{Banerjee06b}
and~\cite{Banerjee07a}.

We find that Bonnor-Ebert models for collapsing magnetized cores
result in enormous accretion rates and outflows that are driven by
toroidal magnetic pressure that is built up in the massive disk. When
filamentary structure is added to this picture, infall becomes even
stronger as gas is funneled from large scales into the filament, and
then down onto the forming disk.

\section{High accretion rates during the collapse phase}
\label{sec:accretion}

\bef
\showtwo{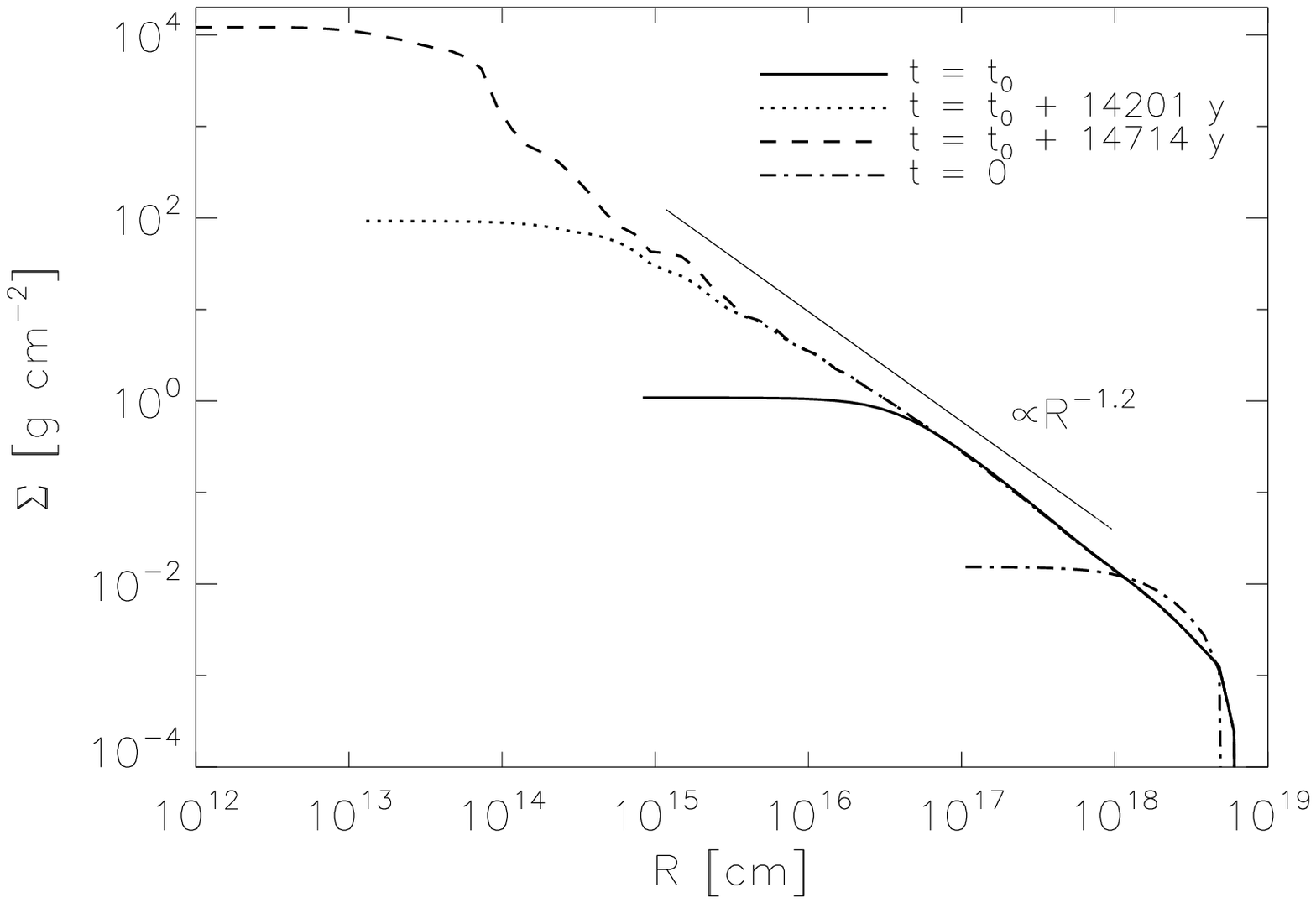}
        {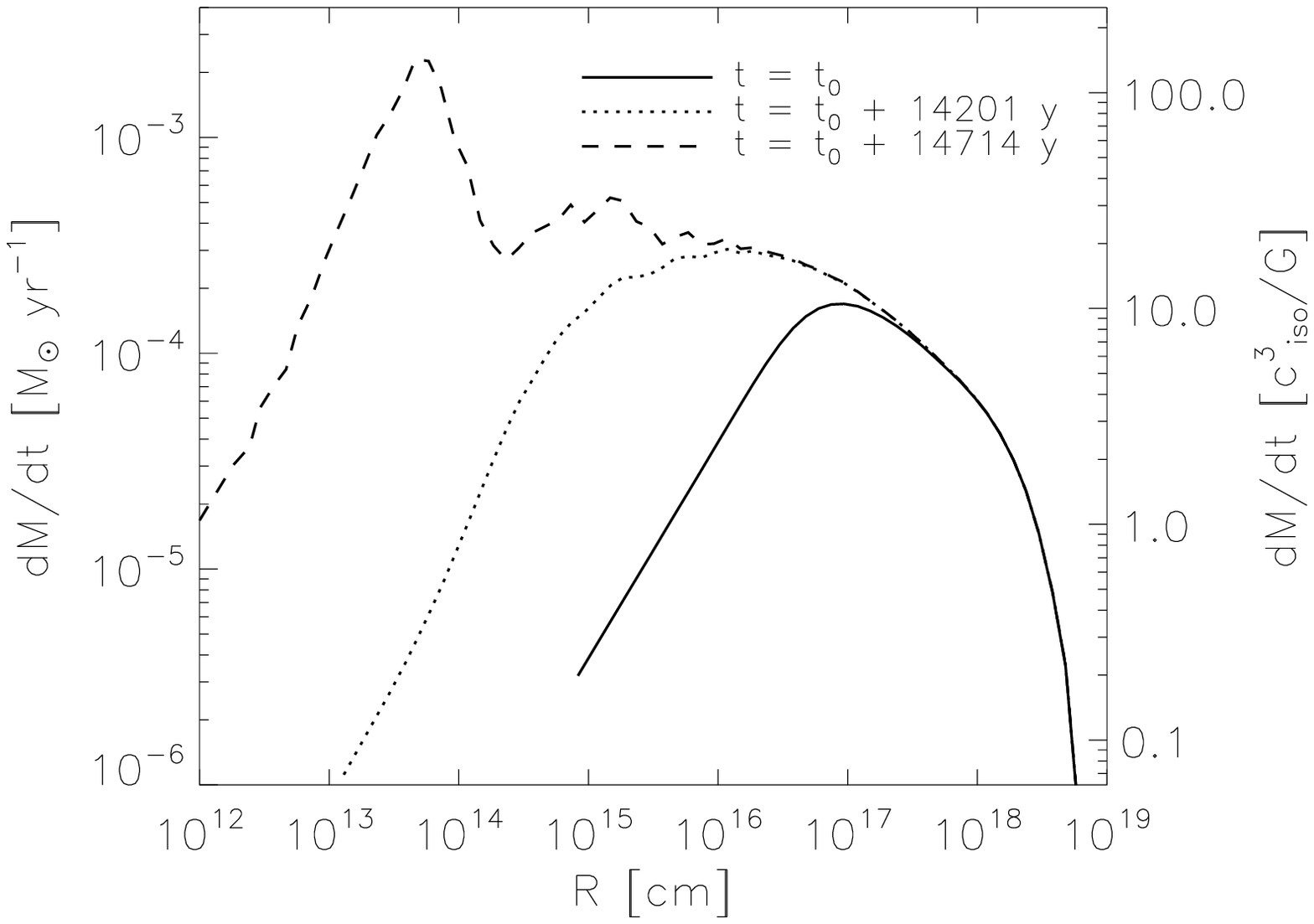}
\caption{Shows the time evolution of the column density and mass
  accretion of a collapsing magnetized core.  We assume the beginning of the
  collapsing phase at $t = t_0$ when the column density reaches
  $\Sigma_{\sm{core}} \approx 1 \, \g \, \cm^{-2}$ and the pressure is
  $P/k_B \approx 10^8 \, \K \, \cm^{-3}$. At $t = t_0$ the mass density in the
  core is $\rho_{\sm{core}} = 2.44\times 10^{-17} \, \g \, \cm^{-3}$
  corresponding to a free fall time of $t_{\sm{ff}} = 1.35\times 10^4
  \, \ys$. At $t = t_0 + 1.1 \, t_{\sm{ff}}$ the core density reached
  already $10^4 \, \g \, \cm^{-2}$. The initial profile of our
  simulation is marked by $t = 0$. [Adapted from~\cite{Banerjee07a}]}
\label{fig:evol_dens}
\eef
 
The collapse of spherical cloud cores has been studied by many
authors, both analytically and numerically \citep[see,
e.g.,][]{Larson69, Penston69, Shu77, Hunter77, Whitworth85, Foster93,
Hennebelle03, Banerjee04}. In the case of an initial singular
isothermal sphere (SIS) there exist an elegant self-similar solution
to this problem where an expansion wave travels from the (singular)
center outward with the speed of sound initiating an inside-out
collapse~\citep{Shu77}. But, as pointed out by \citet{Whitworth96} a
SIS configuration is unnatural because -- among other difficulties --
its collapse can not produce binaries \citep[see
also][]{Pringle89}. The collapse of non-singular cores proceeds
differently than singular spheres and has distinguishable
implications: First -- as demonstrated by many authors
\citep[e.g.,][]{Larson69, Penston69, Foster93, Hennebelle03,
Banerjee04} -- the collapse proceeds from {\em outside-in} rather than
from inside-out and the density maintains a flat profile at the core
center, where the core size is of the order of the local Jeans length
at every epoch~\citep[see also][for a summary of analytic
solutions]{Whitworth85}. Second, the radial distribution of the infall
velocity peaks at the edge of the flat density core and falls off
quickly towards the center (e.g., see Fig.~3 in \citealt{Banerjee04}). The
velocity becomes also supersonic as the core density increases and the
size of the flat region shrinks. Recent observations of a pre-Class 0
object show that the collapse proceeds supersonically, strongly
supporting a Larson-Penston-type collapse rather than an expansion
wave-type collapse \citep{Furuya06}.  Third -- and of great importance
-- the mass accretion in the non-selfsimilar case is much higher than
predicted from the selfsimilar collapse of a singular isothermal
sphere. We find that the mass accretion in the early phase of the
collapse is~\citep{Banerjee07a}
\be
   \dot{M} \approx 20 - 100 \, \frac{c^3}{G}
\label{eq:mdot}
\ee
($c$ is the isothermal sound speed and $G$ is Newton's
gravitational constant). Note the selfsimilar SIS collapse gives a
mass accretion of only $0.96 \, c^3 / G$. Typical values of the sound
speed in cold cloud cores ($T \sim 20 \, \K$) are of the order of a
few $10^2 \, \m \, \sec^{-1}$ which gives $c^3/G \sim 10^{-6} -
10^{-5} \Msol \, \yr^{-1}$. Our result from numerical simulations are
in agreement with the early analytic results of \citet{Larson69} and
\cite{Penston69} \citep[see also][]{Hunter77, Whitworth85}.

The remarkable point is that the high accretion rates are achieved
even without initial turbulence and during the isothermal phase of the
collapse. The main reason for the high accretion rate in this
idealized case is the supersonic infall velocity, $v_r$, close to the
peak density. Even moderate Mach numbers, $\Ma$, of $2 - 3$ (the
Larson asymptotic Mach number is $\Ma \sim 3$) enhances the mass
accretion rates relative to $c^3/G$ because
\be
c^3/G \to v_r^3/G = \Ma^3 \, c^3/G
\label{eq:accretion}
\ee
in the supersonic limit. Additionally, the core is continually
embedded in a high pressure environment.

Studies of the long term evolution (i.e., beyond one dynamical time) of
a collapsing BE-Sphere by~\citet{Foster93} showed that the mass
accretion is not constant, but decreasing with time after it reached a
peak value. Beyond this turnover point the envelope is drained of gas
resulting in a decreasing accretion rate. This situation might be
different for the collapse of more realistic cloud cores which are not
isolated objects but are surrounded by a clumpy medium whose accretion
might sustain high accretion rates for a longer time.

At the time when the column density reaches $\Sigma \sim 1 \, \g \,
\cm^{-2}$ the mass accretion rate becomes $\sim 10^{-4} \, \Msol \,
\yr^{-1}$ and the surrounding pressure is $10^8 \, \K \, \cm^{-3} \,
k_B$.  These results are in agreement with the turbulent core collapse
model of~\citet{McKee02, McKee03} who showed that this high
pressurized, compact cloud cores accrete gas with accretion rates as
high as $10^{-3} \, \Msol \, \yr^{-1}$. We find that accretion rates
of this order ($\sim 10^{-3} \, \Msol \, \yr^{-1}$) are reached within
only $\sim 14.7\times 10^3 \, \ys$ which corresponds to $1.1$
dynamical times where core column density reaches $\sim 10^4 \, \g \,
\cm^{-2}$ (see Fig.~\ref{fig:evol_dens}). As long as the core stays
isothermal (the efficient cooling regime) the pressure scales with the
column density as $P \propto \Sigma^2$ (i.e., the pressure profile is
close to $R^{-2.4}$) and slightly steeper in the inefficient cooling
regime where the temperature rises during the collapse. The continuous
high external pressure and the supersonic infall velocity maintains
the high mass accretion.

\bef
\showtwo{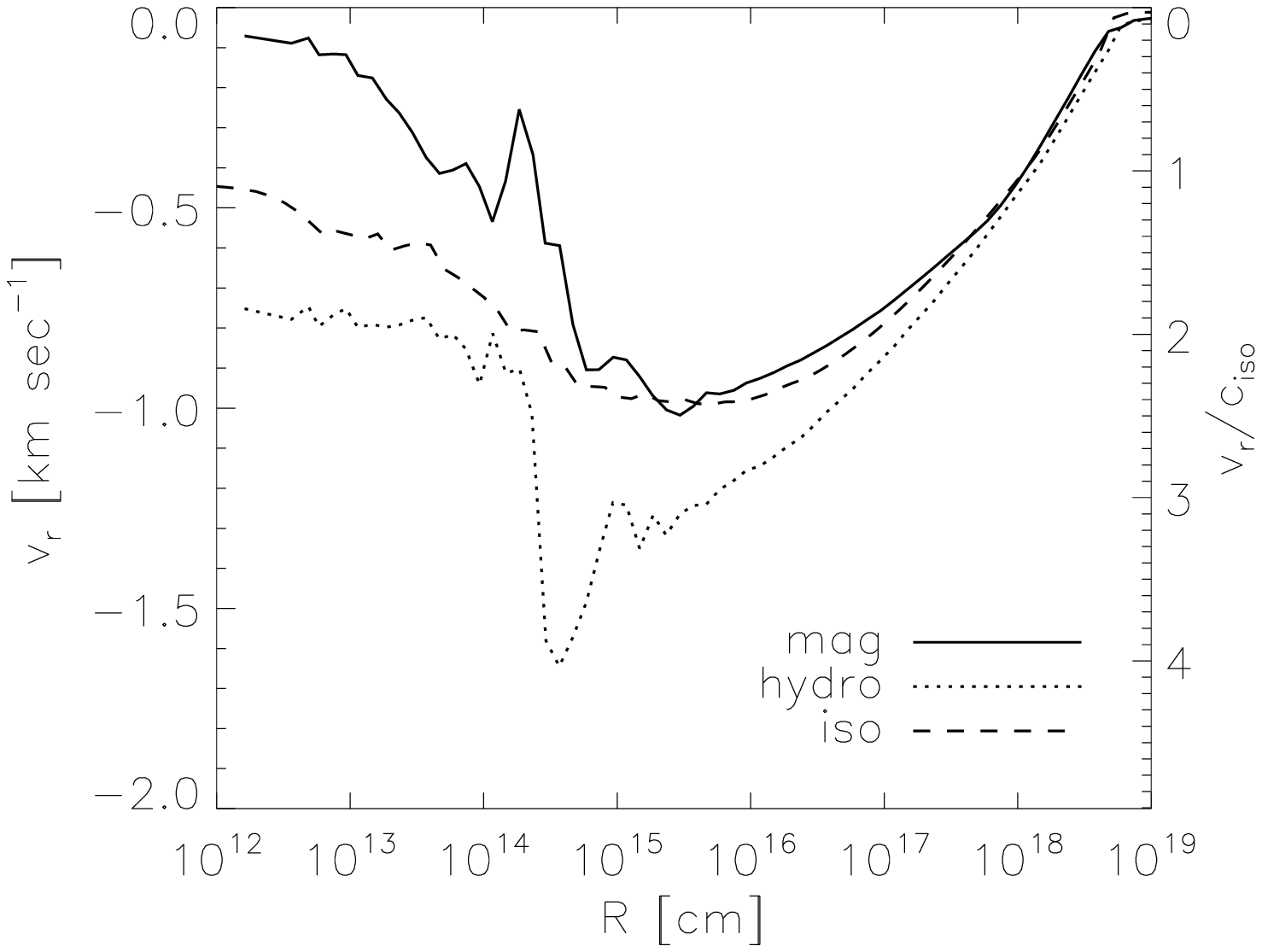}
        {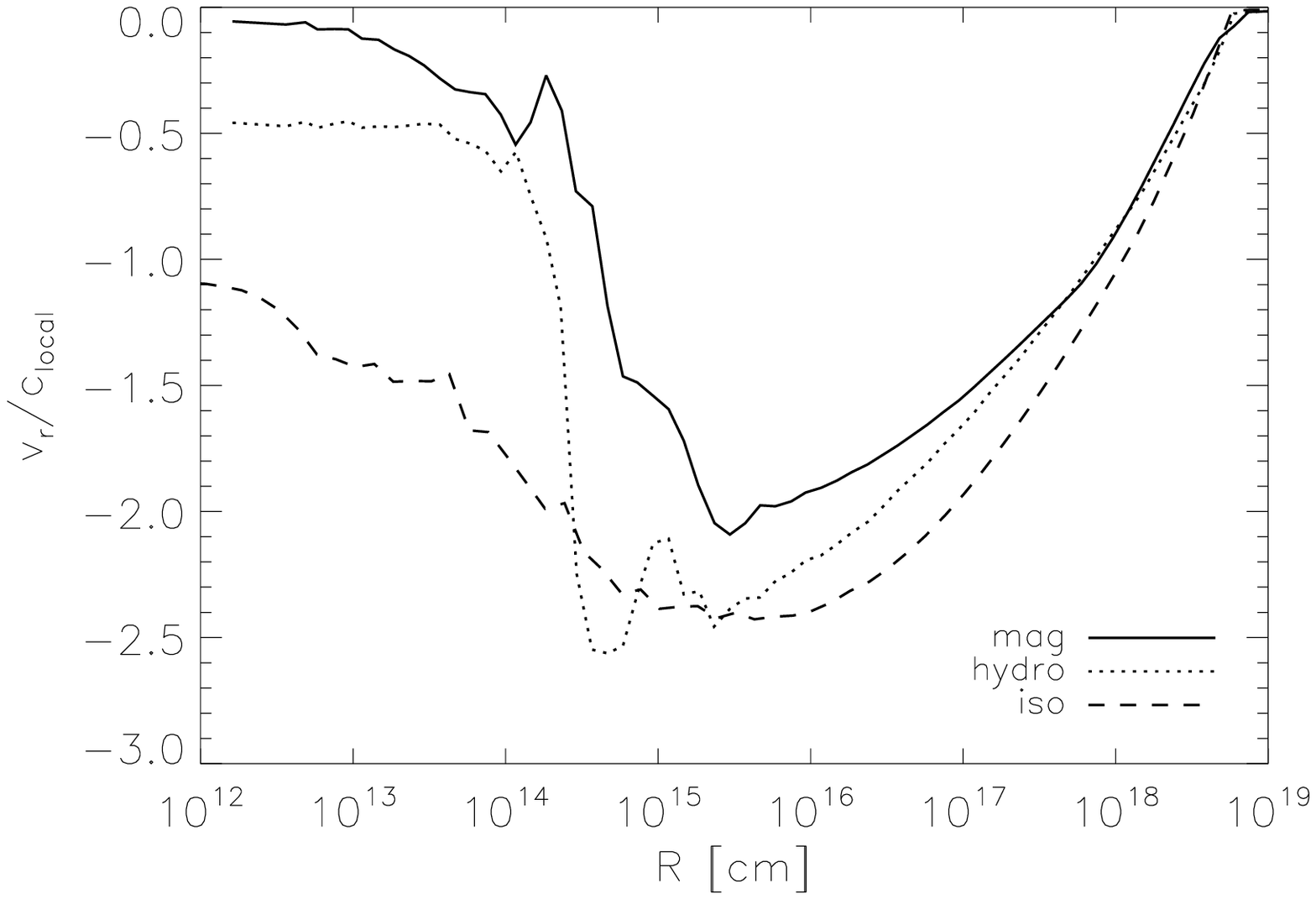}
\caption{Comparison of the radial infall velocity for the three
  different cases, hydromagnetic (mag), pure hydro (hydro), and pure
  isothermal (iso) simulation at the time they reached the same core
  density $\Sigma = 5\times 10^3 \, \g \, \cm^{-2}$. The left panel
  shows the infall velocity in km/sec and compared to the initial
  isothermal sound speed, $c_{\sm{iso}}$, and the right panel shows
  the same quantity measured with the {\em local} sound speed,
  $c_{\sm{local}}$. [Adapted from~\cite{Banerjee07a}]}
\label{fig:vrad}
\eef

In Fig.~\ref{fig:evol_dens} we also show the mass accretion scaled to
the quantity $c_{\sm{iso}}^3/G$ ($c_{\sm{iso}}$ is the {\em initial}
isothermal sound speed) from which one can see that accretion is much
more efficient (by a factor of $\sim 100$) than expected from the
collapse of a SIS. Again, the reason for the high accretion rates is
the supersonic infall of gas. Fig.~\ref{fig:vrad} shows the infall
velocities for a magnetized (mag), pure hydrodynamical case with
cooling (hydro), and an isothermal collapse. Here, the infall
velocities reach up to $1.5 \, \km \, \sec^{-1}$ which corresponds to
four times the {\em initial} sound speed. In all cases the infall
velocities also exceed the {\em local} sound speed where the local
Mach numbers vary between $2$ and $2.5$.

These results show that high mass accretion rates are a natural result
of the early collapse phase of (non-singular) collapsing cloud cores
with flat-topped density profiles. Turbulent driving for the rapid
assembly of massive stars is not a necessary ingredient, but it does
further enhance the accretion rate as shown in analytic
models~\citep{McKee02, McKee03} and even more strongly in fully
realized 3D turbulence~\citep[see, e.g.,][]{Banerjee06b}.

\section{Early outflows during massive star formation}
\label{sec:outflows}

Very little is known observationally about the influence of outflows
on the early assembly of massive stars. On the one hand hand, they
could reduce the mass accretion onto the massive (proto)star if the
outflow carries a substantial mass. On the other hand, early outflows
provide a natural anisotropy of the accreting gas which results in low
density cavities. Such cavities are like funnels out of which the
radiation from the already active star can escape.  Without such a
pressure-release valve, trapped radiative flux would halt the
infall. \cite{Krumholz05} studied the effect of outflow-funnels using
a Monte Carlo radiative-transfer method which shows that radiation
pressure is greatly reduced by radiation escaping the outflow
cavities.

Magnetic fields coupled to the protostellar disk can be the driving
power for such outflows. A variety of self-consistent simulations of
collapsing magnetized cloud cores show that outflows are launched if
the toroidal magnetic field pressure overcomes the gravitational force
\citep[][]{Tomisaka98, Tomisaka02, Matsumoto04, Machida04,
Banerjee06}. Such early-type outflows can be understood in terms of a
growing magnetic tower \citep[][]{LyndenBell03}: The rotating
(proto)disk generates a strong toroidal field component by winding up
the threading field lines. The resulting magnetic pressure is in local
equilibrium with the gravitational force and the ram pressure of the
infalling material. But every new rotation increases the toroidal
field component thereby shifting the equilibrium location (which is
characterized by a shock) towards higher latitudes. The result is an
inflating magnetic bubble in which material is lifted off the disk.

\bef
\showtwo{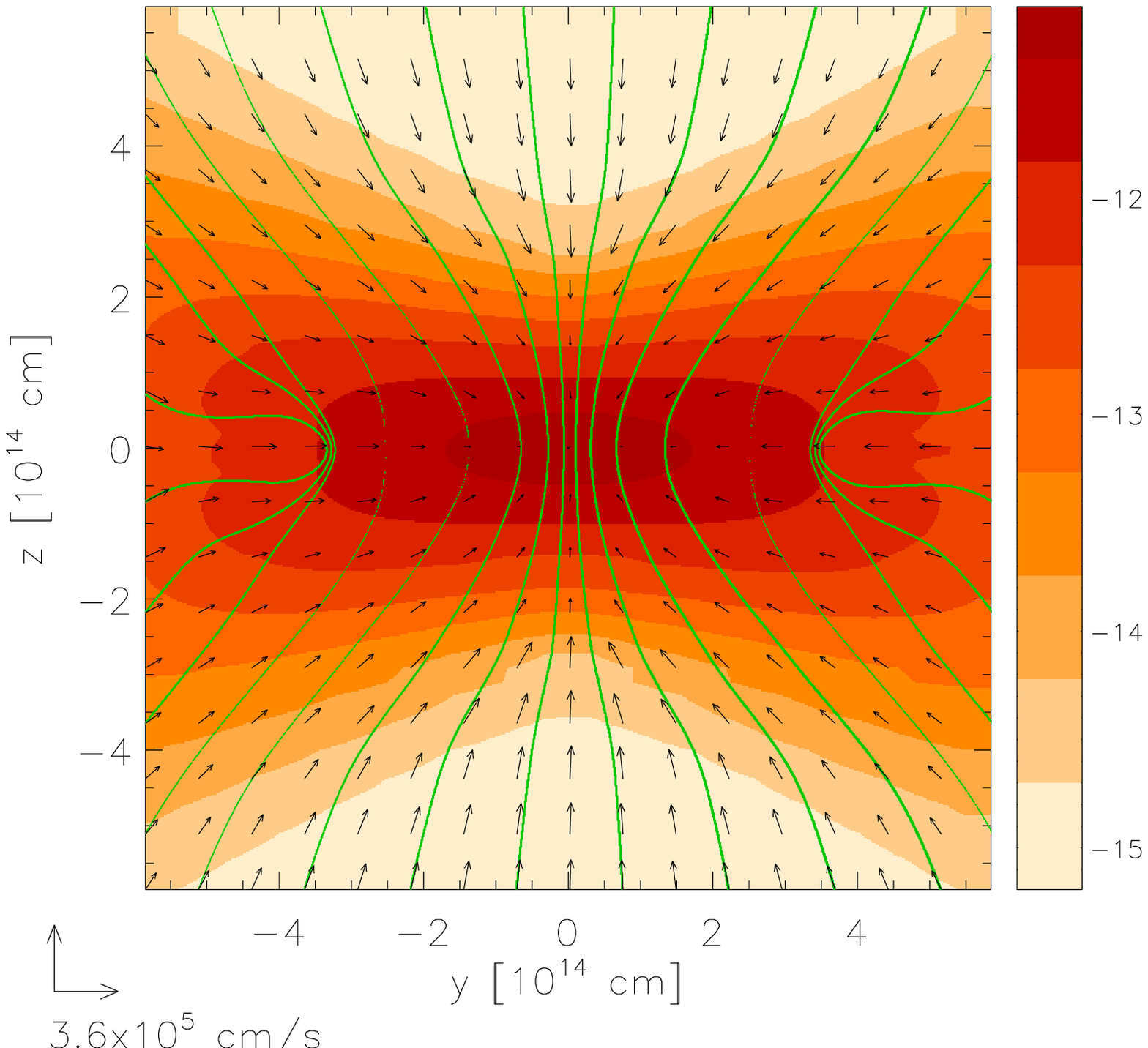}
        {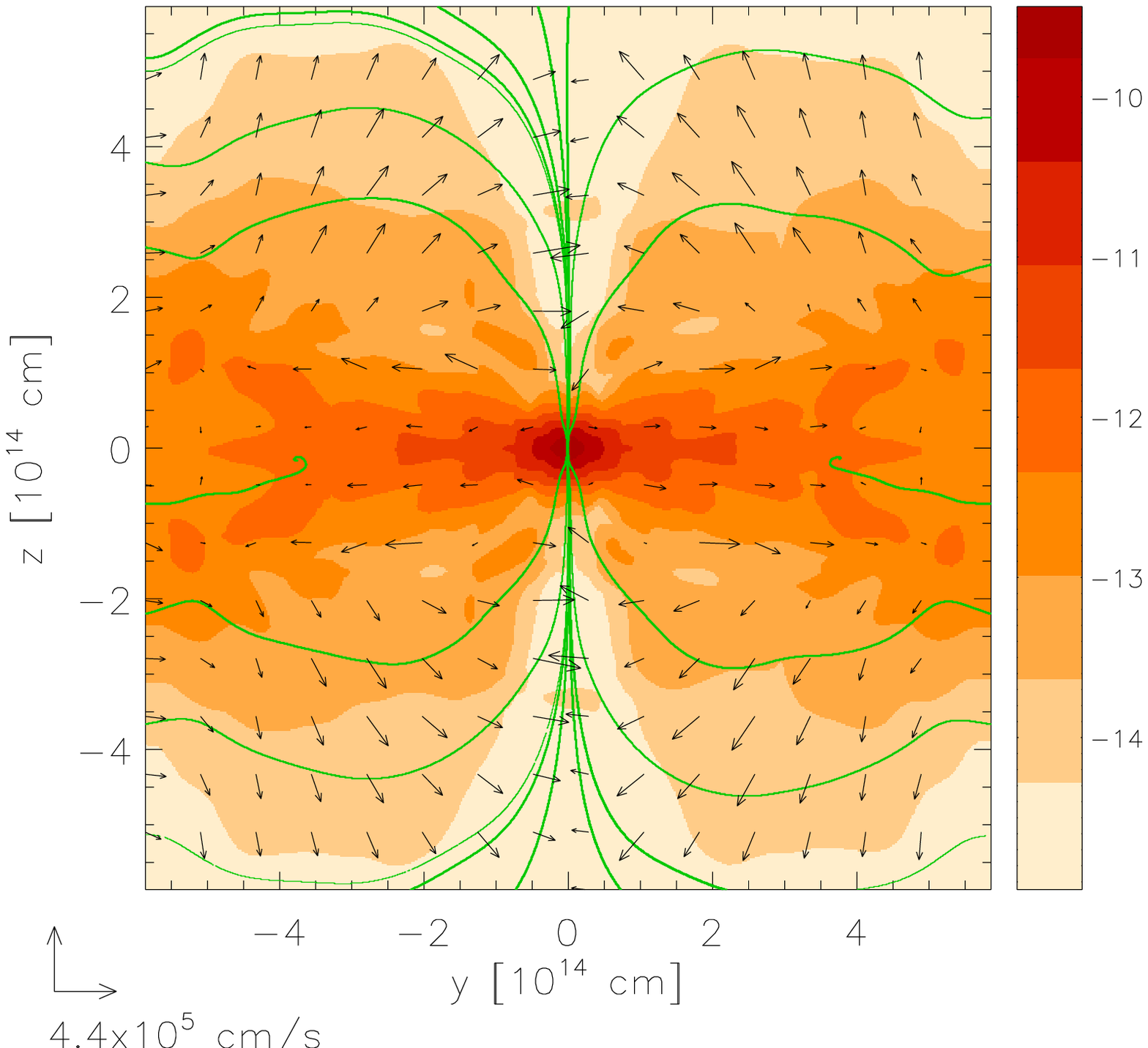}
\caption{Close-up 2D snapshots of the disk region (perpendicular to
  the disk in the $yz$-plane) in the magnetized case. The left panel shows the
  situation at $t = 1.45\times 10^4 \, \ys$ ($1.08 \, t_{\sm{ff}}$)
  into the collapse and before the flow reversal and right panel shows
  the configuration $188 \, \ys$ later when the outflow is clearly
  visible. [Adapted from \cite{Banerjee07a}.]}
\label{fig:small_scale_outflow}
\eef

We explicitly demonstrate the launch of the outflow in the 2D cuts
through our 3D data, shown as the images in
Fig.~\ref{fig:small_scale_outflow} which show the collapse state at
different times.  These close-ups ($\sim 20 \, \AU$) of the early
stage of the outflow launching show the collapsing stage shortly
before the flow reversal (left panel) and the onset of the outflow
(less than $200 \, \ys$ later, right panel).

\section{Massive star assembly in supersonic turbulence}
\label{sec:turbulence}

Our current understanding of star formation is that stars are a
natural consequence of supersonic turbulence within self-gravitating,
molecular clouds \citep[see recent reviews by][]{McKeeOstriker07,
Ballesteros07, Elmegreen04, MacLow04}.

Supersonic turbulence is observed in most if not all giant molecular
clouds (GMCs) and is important because it rapidly sweeps up large
volumes of gas and compresses it into systems of dense filaments.
Observational surveys confirm that filamentary substructure
characterizes the internal organization of molecular clouds on many
scales.  It is clearly evident in the Orion A molecular cloud
\citep[e.g.,][]{Johnstone99} where, in addition to showing the obvious
integral-filament shaped structure, one also sees smaller structures
of $1.3 \, \pc$ in scale.  JCMT submillimeter studies of Orion B
\citep{Mitchell01} reveal a plethora of filamentary structures and
their embedded cores.  This pattern is also seen in the observational
studies by \citet{Fiege04}, wherein a few bright cores are seen to be
embedded in a larger, isolated filamentary structure.  Similar results
are also seen in more embedded regions such as the Lupus 3 cloud where
strong links between filaments and emerging star clusters are observed
\citep{Teixeira05}

\bef
\showone{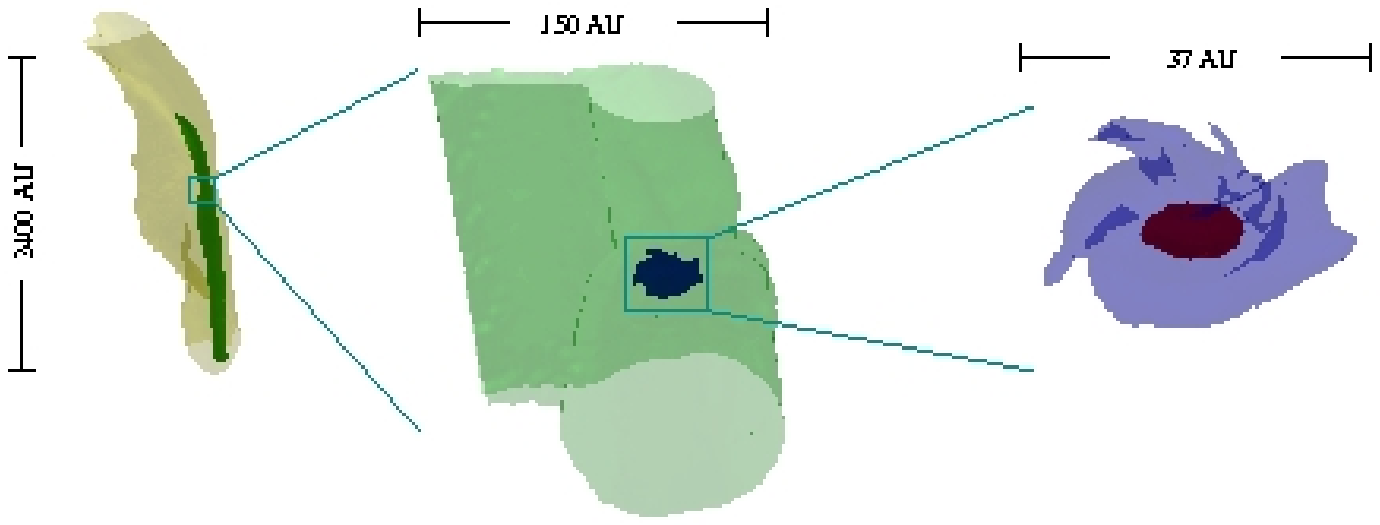}
\caption{Collapse of a elongated turubulent core which is built up in
  a supersonic turbulent molecular cloud~\citep[see][]{Tilley05}.
  Shown are the density isosurfaces at different scales: the
  filamentary structure ($2400 \, \AU$), the protostellar disk within
  the filament ($150 \, \AU$), and the inner region of the disk ($37 \,
  \AU$). [Adapted from \citet{Banerjee06b}.]}
\label{fig:3dstructure}
\eef

Following the collapse of a massive elongated core which is formed
within a supersonic environment (the initial data where taken from one
of the \citet{Tilley05} simulations), we explore the multi-scale
nature of the formation of filamentary large scale structure and how
this controls the formation of protostellar disks and stars in
turbulent cluster-forming clumps.  We follow the filament that forms
the first massive star in our simulation, and find that it has high
density and dynamic pressure that is maintained by continued inflow
and shocking into the filaments.  The collapse of this dense material
along the filament and into the disk and star is the key to
understanding this problem.  Lower mass stars in this picture also
form in filamented structure, but which are less compressed and on
smaller scales.  Our simulations trace how the most massive
star-forming region is assembled and how collapse and the formation of
protostellar disks can occur, by resolving the local Jeans length down
to scales approaching that of the protostar itself.

We find that filaments play a dominant role in controlling the
physics, accretion rate, and angular momentum of the much
smaller-scale accretion disk that forms within such collapsing
structures \citep[see also][]{Balsara01b}. Large-scale filamentary
flows sustain accretion rates that are orders of magnitude greater
than both (i) the naive scalings that are derived from the virial
theorem applied to uniform, 3D media, or even (ii) the collapse of
isolated Bonner-Ebert spheres.  The large-scale structuring of
molecular clouds into filaments therefore has profound effects on the
rate of formation of disks and stars.  

In Figure~\ref{fig:3dstructure} we show the 3D structure of the
collapsing filament at different scales which summarizes our results.
The large-scale filament is about $2600 \, \AU$ in length and one sees
the converging accretion flow on the $150 \, \AU$ scale view presented
in the middle panel.  On this intermediate, $150 \, \AU$ scale, the
disk grows within a filament that is developing out of a sheet-like
structure. The transfer of material from the sheet into the filament
is associated with net angular momentum - and tends to resemble the
growth of a large-scale, spinning vortex.  The disk, forming on yet
smaller scales, acquires this angular momentum from this larger scale
process.  We see the actual disk at this time in the right panel, on a
scale of only $37 \, \AU$ in diameter.

One of the major results of this the filamentary collapse is the very
high accretion rates ($\dot{M} \sim 10^{-2} \, \Msol \, \yr^{-1}$) due
to the supersonic gas flow onto the protostellar disks.  These rates
are $10^3$ times larger than predicted by the collapse of singular
isothermal spheres and exceed the accretion rates necessary to squeeze
the radiation field of the newly born massive star~\citep{Wolfire87}.
We find that a reasonable scaling for filamentary accretion is
$\dot{M} \simeq f \, c^3/G$ where the prefactor should scale as $f
\simeq \Ma^3$, i.e., that one replaces the sound speed in the virial
formula with the free-fall speed.  Even this does not quite account
for the full effect, and there is an additional factor of order a few
that likely is geometric in nature and accounts for filamentary,
rather than spherical infall geometry.
Filamentary structure, therefore, plays a very significant role in star
formation by funneling flows, gathered from large sales, along the
filaments and into the disks.

Our future work will pursue these insights by including the diffuse
radiation from the accretion process and ionization feedback from the
massive star. This will show whether our results are relevant for 
more realistic environments.

\acknowledgements We thank the Kavli Institute for Theoretical
Physics, where we finished this article, for its support, hospitality
and inspiring atmosphere.





\end{document}